\documentclass[11pt,twoside]{article}
\usepackage{asp2004}
\usepackage{psfig}
\usepackage{epsf}
\usepackage{graphics}
\usepackage{lscape}
\def\edcomment#1{\iffalse\marginpar{\raggedright\sl#1\/}\else\relax\fi}
\reversemarginpar

\begin{document}
\title{Populating the Galaxy with close Be binaries}
 \author{P. D. Kiel$^{1}$, J.R. Hurley$^{2}$, J.R. Murray$^{1}$, K. Hayasaki$^{1,3}$}
\affil{$^{1}$Centre for Astrophyscis and Supercomputing, Swinburne University of Technology, Hawthorn, Victoria, 3122, Australia\\
$^{2}$Department of Mathematics, PO Box 28M, Monash University, Victoria 3800, Australia\\
$^{3}$Department of Applied Physics, Graduate School of Engineering, Hokkaido University}


\begin{abstract}
Be/X-ray binaries comprise roughly two-thirds of the 
high-mass X-ray binaries (HMXBs), which is a class of X-ray 
binaries that results from the high mass of the companion or 
donor star ($> 10 M_\odot$). 
Currently the formation and evolution of X-ray producing Be 
binaries is a matter of great debate. 
Modelling of these systems requires knowledge of Be star evolution 
and also consideration of how the evolution changes when the star 
is in close proximity to a companion. 
Within this work we complete a full population synthesis study
of Be binaries for the Galaxy.
The results for the first time match aspects of the observational data,
most notably the observed upper limit to the period distribution. 
We conclude that greater detailed studies on the evolution of Be stars
within X-ray binaries needs to be completed, so that rapid binary 
evolution population synthesis packages may best evolve these systems.
\end{abstract}

\vspace{-0.5cm}
\section{Introduction}

Detailed X-ray surveys have discovered 130 Be/X-ray 
candidates within the Galaxy and Large and Small Magellanic 
Clouds (Raguzova \& Popov, 2005). 
Roughly 50 of these candidates are observed within the Galaxy. 
Only systems with neutron star primaries have been found, though the 
presence of helium star accretors has been suggested in 
some systems (Raguzova, 2001).
We evolve populations of binary stars within the Galaxy following 
both stellar evolution and stellar interactions.
We test how certain assumptions affect the final populations of Be
binaries by varying a number of parameters within different models.
%These calculations and models are outlined in the following sections.

Up to now population synthesis results have not accurately 
matched observed trends. 
One of the most notable trends 
is that of the orbital period cut off, 
where no Be/X-ray system has been observed with 
a period greater than $\sim 600$ days 
(see Figure 3 in Raguzova \& Popov, 2005 as compared to 
Figure 7 of Raguzova, 2001). 
The motivation for this work is, with an up-to-date binary 
evolution and population synthesis package (Hurley, Tout, \& Pols, 2002), 
to attempt to match theoretical calculations with that of observations. 
To do this we calculate relative distributions 
between the differing Be binary types 
-- helium star (He star), white dwarf (WD), neutron star (NS) and black hole 
(BH) companions -- and consider the period distribution of the population. 

\section{Close Be binary evolution \& Population synthesis}

The evolution of close Be binaries is not clear. With this 
in mind we have implemented two different methods 
for rapidly evolving X-ray producing Be binaries.
Firstly we follow Waters et al., (1989) who assumed that the 
X-ray luminosity of Be binaries is produced by wind-fed 
accretion onto the primary star (He star, WD, NS or BH) emanating 
from the circumstellar disk ($\Phi_{Wind}$ method). 
We simply model this as a disk wind with a mass loss rate, 
$\nu_{\dot{m}}$ -- this is in addition to any stellar wind from 
the surface of the Be star. 
Depending on the orbital parameters some fraction of the material may be 
accreted by the primary via Bondi-Hoyle accretion 
(see Hurley, Tout \& Pols 2002). 
We allow our parameter $\nu_{\dot{m}}$ to vary from $10^{-9}$ to 
$10^{-12} M_\odot \, \rm{yr}^{-1}$ for different models (Hayasaki \& 
Okazaki 2004) and any interaction between wind material is ignored. 
%(for He star primaries that may also have a wind). 
Alternatively we model the mass transfer via Roche-lobe over flow 
(RLOF) from the Be disk ($\Phi_{Roche}$ method). 
When a Be star is formed within this method we assume 
that it has a disk extending to a radius of $\mu_{Be} \times R$ 
where $R$ is the radius of the star. 
If the disk is detected to extend beyond the Roche-lobe radius 
of the star then mass-transfer from the Be star is initiated. 
The mass-transfer rate is set by the $\nu_{\dot{m}}$ parameter
and the maximum accretion rate is $10\%$ of this value (Kimitake Hayasaki, 
private communication).
Once a system is evolved, we look for evolutionary phases which fit
our criteria for a Be binary: 
a secondary star spinning at 70\% or greater of its break-up speed  
that is on the MS or first giant branch with mass $\geq 10 M_\odot$,  
and a He star, WD, NS or BH primary which is accreting 
matter by either the RLOF or wind method.

The population synthesis simulations were run using a 
Monte Carlo prescription, a standard initial mass function 
of Kroupa, Tout, \&  Gilmore, (1993) is used separately 
for both primary and secondary masses, with $\alpha = 2.7$.
An initial period function flat in log(P) was also used. 
We evolve primary and secondary masses within a range 
of $5$ to $80 M_\odot$ and $1$ to $80 M_\odot$ respectively. 
The initial period for all models is within a range of $10$ 
to $10\ 000$ days. 
We also assume initially circular binaries with solar metallicities.
Models here vary the parameters $\nu_{\dot{m}}$, 
$\mu_{Be}$ and the method of Be binary evolution ($\Phi_{Roche}$ or 
$\Phi_{Wind}$). 

\section{Results \& Analysis}

Results are given in Table~1. 
The two evolutionary methods do not reflect observations
in that both produce He, WD, NS and BH companions.
However, the two methods do produce a greater relative 
number of Be/NS systems. 
If we calculate an accretion luminosity, $L_x$, and only count systems 
with $L_x > 0.0001 \, L_\odot$ (as opposed to $L_x > 0$, i.e. all systems) 
we find that for the $\Phi_{Wind}$ method the relative number of He to NS 
systems sharply decreases. 
We note that $0.0001 \, L_{\odot}$, or about $10^{29} \,$ergs/s, 
corresponds to the observational lower limit suggested by 
Waters et al., (1989). 
Decreasing $\nu_{\dot{m}}$ below $10^{-9} M_\odot \, {\rm yr}^{-1}$ for the 
$\Phi_{Wind}$ method does not affect the results because the mass-loss rate 
from the wind of the Be star itself will be equal to or greater than this. 
We also see that the choice of $\nu_{\dot{m}}$ has no effect on the 
$\Phi_{Roche}$ numbers however, 
lowering $\mu_{Be}$ does decrease the average X-ray phase life-time.
The $\Phi_{Roche}$ method fails to account for the observed 
period upper limit, where we find that the average period cut 
off for Models A, B and C is only $\sim 70$ days.
Whereas $\Phi_{Wind}$ can roughly simulate this upper limit, with all models
producing a period limit of $\sim 450$ days.
Therefore, we suggest that the $\Phi_{Roche}$ method is flawed 
and some manner of wind accretion is required.

\begin{table}[!ht]
\caption{Models varying different evolutionary parameters.
  Results given show the relative number of systems 
  at a given age of the Galaxy ($13\ 000$ Myr)
  for two lower limit accretion luminosity values.}
\smallskip
\begin{center}
{\small
\begin{tabular}{cccccc}
\tableline
\noalign{\smallskip}
Model & Method & $\nu_{\dot{m}}$ & $\mu_{Be}$ & \multicolumn{2}{c}{Relative numbers; He: WD: NS: BH} \\
      &   &   &   & $L_{x} > 0.0 L_\odot$ & $L_{x} > 0.0001 L_\odot$ \\
\noalign{\smallskip}
\tableline
\noalign{\smallskip}
A & $\Phi_{Roche}$ & $10^{-9}$ & $8$ & $0.25:0.07:1.00:0.06$ & $0.25:0.07:1.00:0.06$ \\
B & $\Phi_{Roche}$ & $10^{-12}$ & $8$ & $0.25:0.07:1.00:0.06$ & $0.25:0.07:1.00:0.06$ \\
C & $\Phi_{Roche}$ & $10^{-9}$ & $4$ & $0.38:0.05:1.00:0.25$ & $0.38:0.05:1.00:0.25$ \\
D & $\Phi_{Wind}$ & $10^{-9}$ & $1$ & $0.98:0.61:1.00:0.04$ & $0.37:0.61:1.00:0.04$ \\
%D & $\Phi_{Wind}$ & $10^{-12}$ & $1$ & $0.98:0.61:1.00:0.04$ & $0.37:0.61:1.00:0.04$ \\
\noalign{\smallskip}
\tableline
\end{tabular}
}
\end{center}
\end{table}

\section{Conclusion}

With the latest rapid binary evolution code we produce a 
distribution of binary systems with the intent on finding 
the relative numbers of Be binaries with 
differing companions in the Galaxy and to match these against 
observations. 
We have produced two different methods for evolving close Be
binaries rapidly, one involves wind accretion while the other 
checks for Roche-lobe overflow from a Be disk. 
We also investigate the effects of varying assumptions 
within Be binary evolution, such as the wind mass loss 
rate from the Be star and the size of the Be disk. 
Although not all of the observational suggestions are simulated
we do --for the first time-- produce a similar period distribution 
cut off to that of observations.
When considering observations many selection 
effects must be taken into account, especially when suggestions 
are made that {\itshape only} Be/NS X-ray systems exits.
In consideration of this and the results found here 
we feel that future work must be
towards placing greater constraints upon the description and 
evolution of Be binaries; modeling of Be disks, 
the details involved in the accretion onto the Be companion 
and what the estimated observed relative and total
numbers of close Be binary systems are within the Galaxy. 

\acknowledgements 
PDK thanks the organizers of the Active OB star 
conference for financial travel support.

\end{document}